\documentclass[aps,twocolumn,amsmath,amssymb,preprintnumbers]{revtex4}
\usepackage{amsmath} \usepackage{amsfonts} \usepackage{amssymb}
\usepackage{bbm}
\usepackage{epsfig}
\usepackage{graphics}
\usepackage{graphicx}
\newcommand{\be}{\begin{equation}}
\newcommand{\ee}{\end{equation}}
\newcommand{\ba}{\begin{eqnarray}}
\newcommand{\ea}{\end{eqnarray}}
\newcommand{\nn}{\nonumber}
\newcommand{\kr}{\rangle}
\newcommand{\kl}{\langle}
\newcommand{\M}{^{(M)}}
\newcommand{\hmn}{^{mn}}
\newcommand{\tmn}{_{mn}}

\newcommand{\x}{\tilde x_}

\newcommand{\f}{{\cal F}}

\newcommand{\h}{{\cal H}}
\newcommand{\zz}{\tilde z^}
\newcommand{\z}{\tilde z_}
\newcommand{\cl}{{\cal L}}
\newcommand{\y}{(\tilde y)}
\newcommand{\mn}{_{\mu\nu}}
\newcommand{\hl}{\hat{\cal L}}

\newcommand{\D}{{\cal D}}

\makeindex

\begin{document}

\title[ ]{Spinor gravity and diffeomorphism invariance on the lattice}

\author{C. Wetterich}
\affiliation{Institut  f\"ur Theoretische Physik\\
Universit\"at Heidelberg\\
Philosophenweg 16, D-69120 Heidelberg}

\begin{abstract}
The key ingredient for lattice regularized quantum gravity is diffeomorphism symmetry. We formulate a lattice functional integral for quantum gravity in terms of fermions. This allows for a diffeomorphism invariant functional measure and avoids  problems of boundedness of the action. We discuss the concept of lattice diffeomorphism invariance. This is realized if the action does not depend on the positioning of abstract lattice points on a continuous manifold. Our formulation of lattice spinor gravity also realizes local Lorentz symmetry. Furthermore, the Lorentz transformations are generalized such that the functional integral describes simultaneously euclidean and Minkowski signature. The difference between space and time arises as a dynamical effect due to the expectation value of a collective metric field. The quantum effective action for the metric is diffeomorphism invariant. Realistic gravity can be obtained if this effective action admits a derivative expansion for long wavelengths.
\end{abstract}

\maketitle

\section{Introduction}
\label{Introduction}

The conceptual unification of general relativity and quantum theory is one of the central goals of theoretical physics. The aim of the present work is an approach to a consistent and mathematically well defined formulation of quantum gravity within the general framework of quantum field theory. It is based on the formulation of a functional integral which is regularized on a lattice of space-time points. As long as the number of space-time points remains finite we deal with a finite number of degrees of freedom. Then all operations for the functional integral are well defined. The continuum limit of infinite volume or vanishing lattice distance is taken at the end.

The central ingredient for general relativity is diffeomorphism symmetry which accounts for the invariance of the formulation under general coordinate transformations. In the presence of fermions this has to be supplemented by local Lorentz symmetry, with Lorentz transformations acting on spinors and the vierbein. In our view the implementation of these symmetries is crucial, much more basic than the choice of particular degrees of freedom as the metric, the vierbein, spinors, or geometrical objects. In a diffeomorphism invariant quantum field theory a metric will be induced as a collective degree of freedom even if it is not present as a fundamental degree of freedom in the formulation of the functional integral. If the quantum effective action for the metric admits a derivative expansion for long wavelengths only a few terms matter for the long distance physics. The possible terms are strongly restricted by diffeomorphism symmetry. The first two are a cosmological constant term with zero derivatives and an Einstein-Hilbert term involving the curvature scalar $R$ with two derivatives. For a small enough value of the cosmological constant this yields a realistic theory for gravity. This holds even if the strict derivative expansion breaks down in higher orders, for example by terms involving $R^2\ln R$. 

We require the following six criteria for a quantum field theory for quantum gravity:

\begin{itemize}
\item [(1)] For a finite number of lattice points the functional integral is well defined.
\item [(2)] The lattice action and functional measure are invariant under lattice diffeomorphisms. 
\item [(3)] A continuum limit exists where lattice diffeomorphism invariance turns into the continuous diffeomorphism symmetry.
\item [(4)] The lattice theory is invariant under local Lorentz transformations. This symmetry is then preserved in the continuum limit. 
\item [(5)] The continuum limit describes some massless (or very light) degrees of freedom. It comprises gravitational interactions. 
\item [(6)] A derivative expansion gives a reasonable approximation for the gravitational degrees of freedom at long wavelengths.
\end{itemize}

\medskip\noindent
A model obeying these criteria realizes a consistent quantum gravity. It describes realistic gravitational interactions if the cosmological constant is small enough, and if there are no additional massless gravitational degrees of freedom beyond the metric or vierbein which induce observable effects like torsion. 

\section{Spinors as fundamental degrees of freedom}
\label{Spinors as fundamental degrees}

For the formulation of a functional integral we first have to decide for the ``fundamental degrees of freedom'' which appear in the action and the functional measure. There are various proposals for lattice formulations, based on spinors \cite{LD4}, \cite{CWLD}, \cite{LD5}, non-linear $\sigma$-models \cite{CWLD}, vierbein and spin connection, the length of edges of simplices in Regge gravity \cite{Ha} or other objects of lattice geometry \cite{Am}, \cite{LD3}.

The metric is a central object for general relativity. It would therefore seem natural to use directly the metric field variables for the formulation of the functional integral. Such an approach encounters, however, two major difficulties. The first is the difficulty to find a functional measure on the lattice that respects diffeomorphism invariance. The second is the problem of boundedness of the action. Depending on the metric configuration the curvature scalar can be positive or negative. Also the determinant of the metric has no definite sign unless one imposes a non-linear signature-constraint. These problems carry over to a formulation with the vierbein as fundamental degree of freedom. Now it is the determinant $e$ of the vierbein that is not positive definite. This determinant appears as a multiplicative factor in the Lagrange density ${\cal L}$, as required by the transformation of ${\cal L}$ as a scalar density under diffeomorphisms. With action $S=\int_x{\cal L}=\int_xeL$ and $L$ a diffeomorphism invariant one can construct configurations of the vierbein with arbitrarily large positive or negative values of $S$. 

In Regge gravity the metric as basic degree of freedom is replaced by the lengths of edges of simplices from which a metric can be computed. One can formulate an invariant lattice functional measure. Despite some remarkable achievements it remains unclear at the present stage if the problem of ``boundedness of the action'' can be overcome such that a satisfactory continuum limit is reached. 

Spinors as basic degrees of freedom avoid both problems. They transform as scalars under general coordinate transformations and the formulation of an invariant functional measure therefore poses no problem. Since spinors are Grassmann variables the functional integral becomes a Grassmann functional integral. Grassmann integrals are always well defined for a finite number of Grassmann variables such that the problem of boundedness of the action is completely absent. Scalar fields with a non-linear constraint (non-linear $\sigma$-models) are also a possibility. The functional measure is trivially diffeomorphism invariant, and the action can become bounded since the constraint forbids arbitrarily large values of the fields.

In this work we concentrate on spinors as basic degrees of freedom. We will discuss a lattice formulation which is close in spirit to spinor gravity as formulated in refs. \cite{HCW,CWSG,CWLL}. The metric and the vierbein arise then as collective fields involving an even number of spinors. First observations that a diffeomorphism invariant action for fermions can be formulated without the use of a metric, and the conjecture that the metric is a composite field, have been made long ago \cite{Aka,Ama,Den}.  (The actual implementation in these early approaches is not always fully consistent - for example the inverse of products of Grassmann variables does not exist.) We build on these ideas, but we propose a different action that implements local Lorentz symmetry.

For continuous spacetime the action of spinor gravity has to be diffeomorphism invariant. A loop expansion for a simple model \cite{HCW,CWSG} has indeed shown that the quantum effective action for the metric can indeed contain Einstein's curvature scalar. Early formulations of spinor gravity as in \cite{HCW,CWSG} exhibit, however, only global and not local Lorentz symmetry. 

A simple geometrical quantity that can be constructed from two spinors is the vierbein bilinear \cite{Aka},
\be\label{AA}
\tilde E^m_\mu=i\bar\psi\gamma^m\partial_\mu\psi,
\ee
with $\gamma^m$ the Dirac matrices. The spinors $\psi$ are scalars with respect to diffeomorphisms, such that $\tilde E^m_\mu$ is a vector. With respect to global Lorentz transformations the vierbein bilinear also transforms as a vector. We may consider $\tilde E^m_\mu$ as a matrix with first index $\mu$ and second index $m$. Then the determinant $\tilde E=\det(\tilde E^m_\mu)$ transforms as a scalar density under general coordinate transformations and is invariant under global Lorentz transformations. An action
\be\label{AB}
S=\sim \int d^4x \tilde E
\ee
is diffeomorphism symmetric and invariant under global Lorentz transformations. One may try to identify the expectation value of $\tilde E^m_\mu$ with the vierbein. Doing so, one encounters the problem that the action \eqref{AB} is invariant under global but not under local Lorentz transformations \cite{HCW}, \cite{CWSG}. The lack of local Lorentz symmetry leads to additional massless degrees of freedom that are contained in the vierbein, beyond the ones corresponding to the metric. (In case of local Lorentz symmetry these would be gauge degrees of freedom.) The resulting torsion terms in the effective action have been discussed in detail in ref. \cite{CWSG}. It was found that one of the torsion invariants - the only one generated at one loop order - is actually compatible with all present observations, while a second possible invariant is excluded by the tests of general relativity. In the present work we avoid this difficulty by formulating a model with local Lorentz symmetry, with analogies to the higher dimensional model in ref. \cite{CWLL}.

The formulation of a basic theory only involving fermions can be viewed as a possible path towards a unified theory of all interactions. In this case all bosons arise as collective fields - in distinction to supersymmetry where fermions and bosons are on equal footing. Gravitons, photons, gluons, $W$- and $Z$-bosons and the Higgs scalar are all composite. Only at length scales large compared to the Planck length they look fundamental, similar to the bosonic hydrogen atom at length scales large compared to Bohr radius. Realizing massless bosonic bound states in a purely fermionic theory is quite common in other physical systems. For example, a Nambu-Jona-Lasinio model \cite{NJL} with spontaneous symmetry breaking of the global chiral symmetry leads to massless pions if the chiral symmetry is exact. Many ``massless'' bosonic excitations are known in condensed matter physics, as spin waves for an antiferromagnet in case of spontaneous breaking of the continuous spin-rotation symmetry. Usually, there is some physical reason for the presence of massless bosons, as the Goldstone theorem for spontaneously broken continuous symmetries. In our case a massless graviton is related to the spontaneous breaking of diffeomorphism symmetry by the selection of a particular metric for the ``ground state'' or cosmological solution. 

In the present work we concentrate on gravitational degrees of freedom for the composite bosons. Gauge bosons or scalars as the Higgs-doublet can arise either directly in a four-dimensional formulation as suitable fermion bilinears, or by dimensional reduction of a higher-dimensional theory of gravity. In our present formulation we find indeed additional symmetries. For two flavors of fermions the continuum limit exhibits an $SU(2)_L\times SU(2)_R$ gauge symmetry. 

Most important, the local Lorentz-transformations of the group $SO(1,3)$ are extended to complex transformation parameters realizing the group $SO(4,{\mathbbm C})$, which also includes the euclidean rotation symmetry $SO(4)$. No signature for space and time are singled out in the basic formulation - both appear on completely equal footing. The difference in signature between space and time arises as a dynamical effect through expectation values of composite fields \cite{CWLL}. 

In the first part of this work we will discuss the continuum action for our
model of spinor gravity. It exhibits diffeomorphism symmetry and local
Lorentz symmetry. We proceed to the lattice formulation in the second part.
The third part investigates the issue of lattice diffeomorphism
invariance. The fourth part addresses the emergence of geometry from lattice spinor gravity. We describe the diffeomorphism invariant quantum effective action for the collective metric field. Our treatment will be largely based on refs. \cite{LD4}, \cite{CWLD}.

\section{Action and functional integral}
\label{Action and functional integral}

Let us explore a setting with $16$ Grassmann variables $\psi^a_\gamma$ at every spacetime point $x$, $\gamma=1\dots 8,~a=1,2$. Here $\gamma$ denotes the eight real variables of a complex four-component Dirac spinor and $a$ is a flavor index for two species of Dirac fermions. The coordinates $x$ parametrize four real numbers, i.e. $x^\mu=(x^0,x^1,x^2,x^3)$. In the lattice formulation these numbers are discrete. We will later associate $t=x^0$ with a time coordinate, and $x^k,~k=1,2,3$, with space coordinates. There is, however, a priori no difference between time and space coordinates. 
The partition function $Z$ is defined as
\ba\label{1A}
Z&=&\int {\cal D}\psi g_f\exp (-S)g_{in},\nn\\
\int {\cal D}\psi&=&\prod_x \prod^{2}_{a=1}
\big\{ \int d\psi^a_1(x)\dots \int d\psi^a_8(x)\big\}.
\ea
For discrete spacetime points on a lattice the Grassmann functional integral \eqref{1A} is well defined mathematically. We assume that the time coordinate $x^0=t$ obeys $t_{in}\leq t\leq t_f$. The boundary term $g_{in}$ is a Grassmann element constructed from $\psi^a_\gamma(t_{in},\vec x)$, while $g_f$ involves terms with powers of $\psi^a_\gamma(t_f,\vec x)$, were $\vec x=(x^1,x^2,x^3)$. If $S$ as well as $g_{in}$ and $g_f$ are elements of a real Grassmann algebra the partition function is real. We may restrict the range of the space coordinates or use a torus $T^3$ instead of ${\mathbbm R}^3$. For a discrete spacetime lattice the number of Grassmann variables is then finite. 

Observables ${\cal A}$ will be represented as Grassmann elements constructed from $\psi^a_\gamma(x)$. We will consider only bosonic observables that involve an even number of Grassmann variables. Their expectation value is defined as
\be\label{2A}
\kl {\cal A}\kr=Z^{-1}\int{\cal D}\psi g_f{\cal A}\exp (-S)g_{in}.
\ee
``Real observables'' are elements of a real Grassmann algebra, i.e. they are sums of powers of $\psi^a_\gamma(x)$ with real coefficients. For real $S,g_{in}$ and $g_f$ all real observables have real expectation values. We will take the continuum limit of vanishing lattice distance at the end. Physical observables are those that have a finite continuum limit. 

For the formulation of the action we will first investigate the continuum limit with $x\in {\mathbbm R}^4$. For this purpose we will work with complex Grassmann variables $\varphi^a_\alpha,\alpha=1\dots 4$, 
\be\label{AA1}
\varphi^a_\alpha(x)=\psi^a_\alpha(x)+i\psi^a_{\alpha+4}(x),
\ee
with $\alpha$ the ``Dirac index'' and $a$ the ``flavor index''. We propose an action which involves twelve Grassmann variables and realizes diffeomorphism symmetry and local $SO(4, {\mathbbm C})$ symmetry 
\ba\label{A1a}
S&=&\alpha\int d^4x\varphi^{a_1}_{\alpha_1}\dots\varphi^{a_8}_{\alpha_8}
\epsilon^{\mu_1\mu_2\mu_3\mu_4}\\
&&\times J^{a_1\dots a_8b_1\dots b_4}_{\alpha_1\dots\alpha_8\beta_1\dots\beta_4}
\partial_{\mu_1}\varphi^{b_1}_{\beta_1}\partial_{\mu_2}\varphi_{\beta_2}^{b_2}\partial_{\mu_3}
\varphi_{\beta_3}^{b_3}\partial_{\mu_4}\varphi^{b_4}_{\beta_4}+c.c.,\nn
\ea
where we sum over repeated indices. The choice of $J$ is dictated by the requirement of Lorentz symmetry and will be discussed in the following chapters. The complex conjugation $c.c.$ replaces $\alpha\to\alpha^*,J\to J^*$ and $\varphi_\alpha(x)\to\varphi^*_\alpha(x)=\psi_\alpha(x)-i\psi_{\alpha+4}(x)$, such that $S^*=S$. In terms of the Grassmann variables $\psi^a_\gamma(x)$ the action $S$ as well as $\exp(-S)$ are elements of a real Grassmann algebra. 

Invariance of the action under general coordinate transformations follows from the use of the totally antisymmetric product of four derivatives $\partial_\mu=\partial/\partial x^\mu$. Indeed, with respect to diffeomorphisms $\varphi(x)$ transforms as a scalar, and $\partial_\mu\varphi(x)$ as a vector. The particular contraction with the totally antisymmetric tensor $\epsilon^{\mu_1\mu_2\mu_3\mu_4},\epsilon^{0123}=1$, allows for a realization of diffeomorphism symmetry without the use of a metric.

\section{Generalized Lorentz transformations}
\label{Generalized Lorentz transformations}

We want to construct an action that is invariant under local generalized Lorentz transformations. Thus the tensor $J^{a_1\dots a_8b_1\dots b_4}_{\alpha_1\dots \alpha_8\beta_1\dots\beta_4}$ must be invariant under global $SO(4,{\mathbbm C})$ transformations. We will often use double indices $\epsilon=(\alpha,a)$ or $\eta=(\beta,b)$, $\epsilon, \eta=1\dots 8$. The tensor $J_{\epsilon_1\dots \epsilon_8\eta_1\dots\eta_4}$ is totally antisymmetric in the first eight indices $\epsilon_1\dots \epsilon_8$, and totally symmetric in the last four indices $\eta_1\dots\eta_4$. This follows from the anticommuting properties of the Grassmann variables $\varphi_\epsilon\varphi_\eta=-\varphi_\eta\varphi_\epsilon$. We will see that for any $J$ invariant under global $SO(4,{\mathbbm C})$ transformations the action \eqref{A1a} is also invariant under local $SO(4, {\mathbbm C})$ transformations.

Local $SO(4,{\mathbbm C})$ transformations act infinitesimally as
\be\label{A2}
\delta\varphi^a_\alpha(x)=-\frac12\epsilon_{mn}(x)(\Sigma^{mn}_E)_{\alpha\beta}\varphi^a_\beta(x),
\ee
with arbitrary complex parameters $\epsilon_{mn}(x)=-\epsilon_{nm}(x),m=0,1,2,3$. The complex $4\times 4$ matrices $\Sigma^{mn}_E$ are associated to the generators of $SO(4)$ in the (reducible) four-component spinor representation. They can be obtained from the euclidean Dirac matrices
\ba\label{A3}
\Sigma^{mn}_E=-\frac14[\gamma^m_E,\gamma^n_E]~,~\{\gamma^m_E,\gamma^n_E\}=2\delta^{mn}.
\ea
Subgroups of $SO(4,{\mathbbm C})$ with different signatures obtain by appropriate choices of $\epsilon_{mn}$. Real parameters  $\epsilon_{mn}$ correspond to euclidean rotations $SO(4)$. Taking $\epsilon_{kl}, k,l=1,2,3$ real, and $\epsilon_{0k}=-i\epsilon^{(M)}_{0k}$ with real $\epsilon^{(M)}_{0k}$, realizes the Lorentz transformations $SO(1,3)$. The Lorentz transformations can be written equivalently with six real transformation parameters $\epsilon^{(M)}_{mn}~,~\epsilon\M_{kl}=\epsilon_{kl}$, using Lorentz-generators $\Sigma\hmn_M$ and signature $\eta\hmn=diag(-1,1,1,1)$, 
\be\label{3A}
\delta\varphi=-\frac12\epsilon\M\tmn\Sigma\hmn_M\varphi ,
\ee
with
\be\label{A4}
\Sigma\hmn_M=-\frac14[\gamma^m_M,\gamma^n_M]~,~\{\gamma^m_M,\gamma^n_M\}=\eta^{mn}.
\ee
The euclidean and Minkowski Dirac matrices are related by $\gamma^0_M=-i\gamma^0_E,\gamma^k_M=\gamma^k_E$. 

The transformation of a derivative involves an inhomogeneous part 
\be\label{A5a}
\delta\partial_\mu\varphi_\beta=-\frac12\epsilon\tmn(\Sigma\hmn\partial_\mu
\varphi)_\beta-\frac12\partial_\mu\epsilon\tmn(\Sigma\hmn\varphi)_\beta,
\ee
with $\Sigma\hmn=\Sigma\hmn_E~,~\gamma^m=\gamma^m_E$. The first ``homogeneous term'' $\sim \partial_\mu\varphi$ transforms as $\varphi_\beta$. Thus an invariant tensor $J$ guarantees an invariant action if the second term in eq. \eqref{A5a} does not contribute to $\delta S$. Contributions of the second ``inhomogeneous term'' to the  variation of the action $\delta S$ involve at least nine spinors at the same position $x$, i.e. $(\Sigma\hmn\varphi)^b_\beta(x)\varphi^{a_1}_{\alpha_1}(x)\dots\varphi^{a_8}_{\alpha_8}(x)$. Therefore this inhomogeneous contribution to $\delta S$ vanishes due to the identity $\varphi_\alpha(x)\varphi_\alpha(x)=0$ (no sum here) - at most eight different complex spinors can be placed on a given position $x$. The invariance of $S$ under global $SO(4,{\mathbbm C})$ transformations entails the invariance under local $SO(4,{\mathbbm C})$ transformations. We have constructed in ref. \cite{CWLL} a model for sixteen dimensional spinor gravity with local $SO(16,{\mathbbm C})$ symmetry. The present four-dimensional model shows analogies to this. 

It is important that all invariants appearing in the action \eqref{A1a} involve either only factors of $\varphi_\alpha=\psi_\alpha+i\psi_{\alpha+4}$ or only factors of $\varphi^*_\alpha =\psi_\alpha-i\psi_{\alpha+4}$. It is possible to construct $SO(1,3)$ invariants which involve both $\varphi$ and $\varphi^*$. Those will not be invariant under $SO(4,{\mathbbm C})$, however. We can also construct invariants involving $\varphi$ and $\varphi^*$ which are invariant under euclidean $SO(4)$ rotations. They will not be invariant under $SO(1,3)$. The only terms which are invariant under both $SO(4)$ {\em and} $SO(1,3)$, and more generally $SO(4,{\mathbbm C})$, are those constructed from $\varphi$ alone or $\varphi^*$ alone, or products of such invariants. (Invariants involving both $\varphi$ and $\varphi^*$ can be constructed as products of invariants involving only $\varphi$ with invariants involving only $\varphi^*$.) We conclude that for a suitable $SO(4,{\mathbbm C})$-invariant tensor $J$ the action has the symmetries required for a realistic theory of gravity for fermions, namely diffeomorphism symmetry and local $SO(1,3)$ Lorentz symmetry. No signature and no metric are introduced at this stage, such that there is no difference between time and space \cite{CWLL}.

\section{Lorentz invariant spinor bilinears}
\label{Lorentz invariant spinor bilinears}

We next want to construct the $SO(4,{\mathbbm C})$ invariant tensor $J$ in eq. \eqref{A1a}. We do this in steps by discussing first simpler invariants, out of which we will compose $J$. Our model with two flavors allows us to construct two symmetric invariants with two Dirac indices
\be\label{W3a}
S^\pm_{\eta_1\eta_2}=(S^\pm)^{b_1b_2}_{\beta_1\beta_2}
=\mp(C_\pm)_{\beta_1\beta_2}(\tau_2)^{b_1b_2},
\ee
where $\tau_k$ denotes the Pauli matrices. The invariant tensors $C_\pm$ are antisymmetric \cite{CWS}
\be\label{SD}
(C_\pm)_{\beta_2\beta_1}=-(C_\pm)_{\beta_1\beta_2},
\ee
such that $S^\pm$ is symmetric under the exchange $(\beta_1,b_1)\leftrightarrow(\beta_2,b_2)$, or, in terms of the double index $\eta=(\beta,b)$, 
\be\label{SE}
S^\pm_{\eta_2\eta_1}=S^\pm_{\eta_1\eta_2}.
\ee

The $SO(4,{\mathbbm C})$-invariants $C_\pm$ can best be understood in terms of Weyl spinors. The matrix 
\be\label{37A}
\bar\gamma=-\gamma^0\gamma^1\gamma^2\gamma^3
\ee
commutes with $\Sigma\hmn$ such that the two doublets
\be\label{A25}
\varphi_+=\frac12(1+\bar\gamma)\varphi~,~\varphi_-=\frac12(1-\bar\gamma)\varphi
\ee
correspond to inequivalent two component complex spinor representations (Weyl spinors). We employ here a representation of the Dirac matrices $\gamma^m$ where $\bar\gamma=diag(1,1,-1,-1)$, namely
\be\label{16A}
\gamma^0=\tau_1\otimes 1~,~\gamma^k=\tau_2\otimes \tau_k.
\ee
(The general structure is independent of this choice. Our representation corresponds to the Weyl basis of ref. \cite{CWMS} where details of conventions can be found.) In this representation one has 
\ba\label{W2}
C_+&=&\frac12(C_1+C_2)=\frac12 C_1(1+\bar\gamma)=
\left(\begin{array}{ccc}\tau_2&,&0\\0&,&0\end{array}\right),\nn\\
C_-&=&\frac12(C_1-C_2)=\frac12 C_1(1-\bar\gamma)=
\left(\begin{array}{ccc}0&,&0\\0&,&-\tau_2\end{array}\right).
\ea
such that $\psi^T_\pm C_1=\psi^T_\pm C_\pm=\psi^TC_\pm$.

It is straightforward to construct invariants only involving the two Weyl spinors $\varphi^1_+$ and $\varphi^2_+$ by combining $C_+$ with an appropriate flavor matrix. For this purpose we can restrict the index $\eta$ to the values $1\dots 4$. The action of $SO(4,{\mathbbm C})$ on $\varphi_+$ is given by the subgroup of complexified $SU(2,{\mathbbm C})_+$ transformations. In our basis the generators of $SU(2,{\mathbbm C})_+$ read
\be\label{W6}
\Sigma^{0k}=-\frac i2\tau_k~,~\Sigma^{kl}=\epsilon^{klm}\Sigma^{0m},
\ee
such that $\Sigma^{kl}$ is linearly dependent on $\Sigma^{0k}$. (For $SU(2,{\mathbbm C})_-$ the generators $\Sigma^{kl}$ are identical, while $\Sigma^{0k}=\frac i2\tau_k$. The subgroup of special unitary transformations $SU(2)$ obtains for real transformation parameters, while we consider here arbitrary complex transformation parameters.) 

We observe that we can also consider a group $SU(2,{\mathbbm C})_L$ acting on the flavor indices of $\varphi_+$. With respect to $SU(2,{\mathbbm C})_+\times SU(2,{\mathbbm C})_L$ the four component spinor $\varphi_{+,\eta}~(\eta=1\dots 4)$ transforms as the $(2,2)$ representation. Since the matrix $(\tau_2)^{ab}$ in eq. \eqref{W3a} is invariant under $SU(2,{\mathbbm C})_L$, the invariant $S^+$ is invariant under the group 
\be\label{W7}
SO(4,{\mathbbm C})_+\equiv SU (2,{\mathbbm C})_+\times SU(2,{\mathbbm C})_L. 
\ee
(Here $SO(4,{\mathbbm C})_+$ should be distinguished from the generalized Lorentz transformation since it acts both in the space of Dirac and flavor indices.) With respect to $SO(4,{\mathbbm C})_+$ the two-flavored spinor $\varphi_+$ transforms as a four component vector. The classification of tensors, invariants and symmetries can be directly inferred from the analysis of four-dimensional vectors. Invariants only involving $\varphi_-$ can be constructed in a similar way with $SU(2,{\mathbbm C})_R$ acting on the flavor indices of $\varphi_-$ and $SO(4,{\mathbbm C})_-=SU(2,{\mathbbm C})_-\times SU(2,{\mathbbm C})_R$.

\section{Action with local Lorentz symmetry}
\label{Action with local Lorentz symmetry}

A totally symmetric invariant four index object can be constructed as
\ba\label{W9}
&&L_{\eta_1\eta_2\eta_3\eta_4}=\frac16
(S^+_{\eta_1\eta_2}S^-_{\eta_3\eta_4}+S^+_{\eta_1\eta_3}
S^-_{\eta_2\eta_4}+S^+_{\eta_1\eta_4}S^-_{\eta_2\eta_3}\nn\\
&&\hspace{1.0cm} +S^+_{\eta_3\eta_4}S^-_{\eta_1\eta_2}+S^+_{\eta_2\eta_4}
S^-_{\eta_1\eta_3}+S^+_{\eta_2\eta_3}S^-_{\eta_1\eta_4}).
\ea
The global invariant with four derivatives
\be\label{30D}
D=\epsilon^{\mu_1\mu_2\mu_3\mu_4}
\partial_{\mu_1}\varphi_{\eta_1}\partial_{\mu_2}\varphi_{\eta_2}\partial_{\mu_3}\varphi_{\eta_3}\partial_{\mu_4}\varphi_{\eta_4}
L_{\eta_1\eta_2\eta_3\eta_4}
\ee
involves two Weyl spinors $\varphi_+$ and two Weyl spinors $\varphi_-$. Furthermore, an invariant with eight factors of $\varphi$ involves the totally antisymmetric tensor for the eight values of the double-index $\epsilon$
\ba\label{30E}
A^{(8)}&=&\frac{1}{8!}\epsilon_{\epsilon_1\epsilon_2\dots\epsilon_8}
\varphi_{\epsilon_1}\dots\varphi_{\epsilon_8}\nn\\
&=&\frac{1}{(24)^2}\epsilon_{\alpha_1\alpha_2\alpha_3\alpha_4}
\varphi^1_{\alpha_1}\dots\varphi^1_{\alpha_4}\epsilon_{\beta_1\beta_2\beta_3\beta_4}
\varphi^2_{\beta_1}\dots\varphi^2_{\beta_4}\nn\\
&=&\varphi^1_1\varphi^1_2\varphi^1_3\varphi^1_4
\varphi^2_1\varphi^2_2\varphi^2_3\varphi^2_4.
\ea

An action with local $SO(4,{\mathbbm C})$ symmetry takes the form 
\be\label{30F}
S=\alpha\int d^4xA^{(8)}D+c.c.
\ee
Indeed, the inhomogeneous contribution \eqref{A5a} to the variation of $D(x)$ contains factors $(\Sigma^{mn}\varphi^b)_\beta(x)$. As discussed before, it vanishes when multiplied with $A^{(8)}(x)$, since the Pauli principle $\big(\varphi^a_\alpha(x)\big)^2=0$ admits at most eight factors $\varphi$ for a given $x$. In consequence, the inhomogeneous variation of the action \eqref{30F} vanishes and $S$ is invariant under {\em local } $SO(4,{\mathbbm C})$ transformations. In contrast to $\int d^4xD(x)$ the action $S$ in eq. \eqref{30F} is not a total derivative. Besides local $SO(4,{\mathbbm C})$, it is also invariant under local $SO(4,{\mathbbm C})_F$ gauge transformations, with $SO(4,{\mathbbm C})_F=SU(2,{\mathbbm C})_L\times SU(2,{\mathbbm C})_R$. 

The derivative-invariant $D$ can be written in the form 
\be\label{32A}
D=\epsilon^{\mu_1\mu_2\mu_3\mu_4}D^+_{\mu_1\mu_2}D^-_{\mu_3\mu_4},
\ee
with 
\be\label{32B}
D^\pm_{\mu_1\mu_2}=\partial_{\mu_1}\varphi_{\eta_1}S^\pm_{\eta_1\eta_2}\partial_{\mu_2}\varphi_{\eta_2}.
\ee
Inserting eq. \eqref{32A} into eq. \eqref{30F} we recognize the contraction of four derivatives with the totally antisymmetric $\epsilon$-tensor which explains the invariance of $S$ under diffeomorphisms. Eq. \eqref{32A} shows that $D$ is invariant under the exchange $\varphi_{+,\eta}\leftrightarrow\varphi_{-,\eta}$. The transformation $\varphi\to\gamma^0\varphi$ maps $S^+_{\eta_1\eta_2}\leftrightarrow S^-_{\eta_1\eta_2}$ and therefore $D^+_{\mu_1\mu_2}\leftrightarrow D^-_{\mu_1\mu_2}$, such that again $D$ is invariant. (For our choice $\gamma^0=\tau_1\otimes 1$ the transformation $\varphi\to\gamma^0\varphi$ actually corresponds to $\varphi_{+,\eta}\leftrightarrow \varphi_{-,\eta}$.) We can also decompose
\be\label{32C}
A^{(8)}=A^+A^-,
\ee
with 
\be\label{32D}
A^+=\varphi^1_{+1}\varphi^1_{+2}\varphi^2_{+1}\varphi^2_{+2},
\ee
and similarly for $A^-$. The combinations 
\be\label{32E}
F^\pm_{\mu_1\mu_2}=A^\pm D^\pm_{\mu_1\mu_2}
\ee
are invariant under local $SO(4,{\mathbbm C})\times SO(4,{\mathbbm C})_F$ transformations. They involve six Weyl spinors $\varphi_+$ or six Weyl spinors $\varphi_-$, respectively. The action involves products of $F^+$ and $F^-$, 
\be\label{32F}
S=\alpha\int d^4x\epsilon^{\mu_1\mu_2\mu_3\mu_4}F^+_{\mu_1\mu_2}F^-_{\mu_3\mu_4}+c.c.
\ee

We define the Minkowski action by
\be\label{A23}
S=-iS_M~,~e^{-S}=e^{iS_M},
\ee
which yields the usual ``phase factor'' for the functional integral written in terms of $S_M$. We can define the operation of a transposition as a total reordering of all Grassmann variables. The result of transposition for a product of Grassmann variables depends only on the number of factors $N_\varphi$. For $N_\varphi=2,3$ mod $4$ the transposition results in a minus sign, while for $N_\varphi=4,5$ mod $4$ the product is invariant. In consequence, one finds that $S_M$ is symmetric. With respect to the complex conjugation c.c. used in eq. \eqref{AA1} the Minkowski action is antihermitean. This complex conjugation, which is defined for the Grassmann variables $\psi_\gamma$ by the involution $\psi^a_{\alpha+4}\to-\psi^a_{\alpha+4}$ for $\alpha=1\dots 4$, is, however, not unique. We may define a different conjugation by an involution where the Grassmann variables changing sign are $\psi^1_5,\psi^1_6,\psi^1_7,\psi^1_8,\psi^2_3,\psi^2_4,\psi^2_5$ and $\psi^2_6$. In this case we use the same definition as before for $\varphi^1_\alpha$ and $\varphi^2_1,\varphi^2_2$, but we replace $\varphi^2_3$ and $\varphi^2_4$ by new complex Grassmann variables
\ba\label{AxA}
\xi^2_3&=&\psi^2_7-i\psi^2_3~,~\xi^2_4=\psi^2_8-i\psi^2_4,\nn\\
(\xi^2_3)^*&=&\psi^2_7+i\psi^2_3~,~(\xi^2_4)^*=\psi^2_8+i\psi^2_4.
\ea
The new complex conjugation can be interpreted as a multiplication of c.c. in eq. \eqref{AA1} with the transformation $\varphi^2_-\to-\varphi^2_-$. Expressing the euclidean action in terms of $\varphi^1_\pm,\varphi^2_+$ and $\xi^2_-$ it changes sign under the new complex conjugation. With respect to this conjugation the Minkowski action is real and symmetric and therefore hermitean. We can use the first complex conjugation in order to establish that we work with a real Grassmann algebra, and the second one to define hermiticity of $S_M$ which is related to a unitary time evolution. 

\section{Gauge and discrete symmetries}
\label{Symmetries}

Besides the generalized Lorentz transformations $SO(4,{\mathbbm C})$ the action \eqref{30F}, \eqref{32F} is also invariant under continuous gauge transformations. By the same argument as for local $SO(4,{\mathbbm C})$ symmetry, any global continuous symmetry of the action is also a local symmetry due to the Pauli principle. We have already encountered the symmetry $SU(2,{\mathbbm C})_L$ which transforms
\be\label{S1}
\delta\varphi^a_{+\alpha}(x)=\frac i2\tilde\alpha_{+k}(x)(\tau_k)^{ab}\varphi^b_{+\alpha}(x),
\ee
with three complex parameters $\tilde\alpha_{+k}$, and similar for $SU(2,{\mathbbm C})_R$ acting on $\varphi_-$. For real $\tilde\alpha_{+k}$ these are standard gauge transformations with compact gauge group $SU(2)$. Altogether, we have four $SU(2,{\mathbbm C})$ factors. With respect to $G=SU(2,{\mathbbm C})_+\times SU(2,{\mathbbm C})_-\times SU(2,{\mathbbm C})_L\times SU(2,{\mathbbm C})_R$ the Weyl spinors $\varphi_+$ and $\varphi_-$ transform as $(2,1,2,1)$ and $(1,2,1,2)$, respectively, and the action is invariant.

Discrete symmetries are also a useful tool to characterize the properties of the model. Simple symmetries of the action are $Z_{12}$ phase-transformations or multiplications with $\bar\gamma$ or $\gamma^0$, e.g.
\be\label{51}
\varphi\to \exp (2\pi in/12)\varphi~,~\varphi\to\bar\gamma\varphi~,~\varphi\to \gamma^0\varphi.
\ee
The reflection of the three space coordinates
\be\label{52}
\psi^a_\gamma(x)\to \psi^a_\gamma(Px)~,~P(x^0,x^1,x^2,x^3)=(x^0,-x^1,-x^2,-x^3),
\ee
changes the sign of the action. If this transformation is accompanied by any other discrete transformation which inverts the sign of $S$ the combined transformation amounts to a type of parity symmetry. As an example, we may consider
\be\label{53}
\varphi^1(x)\to\gamma^0\varphi^1(x)~,~\varphi^2(x)\to\gamma^0\bar\gamma\varphi^2(x).
\ee
Time reflection symmetry can be obtained in a similar way by combining $\psi^a_\gamma(x)\to\psi^a_\gamma(-Px)$ with a suitable transformation that changes the sign of $S$, as for eq. \eqref{53}. Reflections of an even number of coordinates, including the simultaneous space and time reflections, $\psi^a_\gamma(x)\to\psi^a_\gamma(-x)$, leave the action invariant.

\section{Discretization}
\label{Discretization}

In the second part we formulate a regularized version of the functional integral \eqref{1A}. For this purpose we will use a lattice of space-time points. We recall that the action \eqref{30F} is invariant under $SO(4)$ and $SO(1,3)$ transformations and does not involve any metric. The regularization will therefore be valid simultaneously for a Minkowski and a euclidean theory.

Let us consider a four-dimensional hypercubic lattice with lattice distance $\Delta$. We distinguish between the ``even sublattice'' of points $y^\mu=\tilde y^\mu\Delta$, $\tilde y^\mu$ integer, $\Sigma_\mu\tilde y^\mu$ even, and the ``odd sublattice'' $z^\mu=\tilde z^\mu\Delta~,~\tilde z^\mu$ integer, $\Sigma_\mu\tilde z^\mu$ odd. The odd sublattice is considered as the fundamental lattice, and we associate to each position $z^\mu$ the $16$ (``real'') Grassmann variables $\psi^a_\gamma(z)$, or their complex counterpart $\varphi^a_\alpha(z)$. (We use here $z$ instead of $x$ in order to make the difference between lattice coordinates and continuum coordinates more visible.) The functional measure \eqref{1A} is invariant under local $SO(4,{\mathbbm C})$ transformations since it can be written as a product of invariants of the type $A_+,A_-$ in eq. \eqref{32D}  and their complex conjugate for every $z$. It is also invariant under local $SU(2,{\mathbbm C})_L\times SU(2,{\mathbb C)})_R$ gauge transformations. 

We write the action as a sum over local terms or Lagrangians ${\cal L}(y)$, 
\be\label{L1}
S=\tilde\alpha\sum_y{\cal L}(y)+c.c.
\ee
Here $y^\mu$ denotes a position on the even sublattice or ``dual lattice''. It has eight nearest neighbors on the fundamental lattice, with distance $\Delta$ from $y$. To each point $y$ we associate a ``cell'' of those eight points $\tilde x_j$ whose $\tilde z$-coordinates are given by 
\be\label{V1}
\tilde z^\mu=\tilde y^\mu\pm (w_\nu)^\mu,
\ee
with $(w_\nu)^\mu=\delta^\mu_\nu$. The Lagrangian ${\cal L}(y)$ is given by a sum of ``hyperloops''. A hyperloop is a product of an even number of Grassmann variables located at positions $\tilde x_j(\tilde y), j=1\dots 8$, within the cell at $\tilde y$. In accordance with eq. \eqref{A1a} we will consider hyperloops with twelve spinors. In a certain sense the hyperloops are a four-dimensional generalization of the plaquettes in lattice gauge theories. 

We want to preserve the local $SO(4,{\mathbbm C})$-symmetry for the lattice regularization of spinor gravity. We therefore employ hyperloops that are invariant under local $SO(4,{\mathbbm C})$ transformations. Local $SO(4,{\mathbbm C})$ symmetry can be implemented by constructing the hyperloops as products of invariant bilinears involving two spinors located at the same position $\tilde z$,
\be\label{L4}
\tilde \h^k_\pm (\tilde z)=\varphi^a_\alpha(\tilde z)(C_\pm)_{\alpha\beta}(\tau_2\tau_k)^{ab}\varphi^b_\beta(\tilde z).
\ee
Since the local $SO(4,{\mathbbm C})$ transformations \eqref{A2} involve the same $\epsilon_{mn}(\tilde z)$ for both spinors the six bilinears $\tilde \h^k_\pm$ are all invariant. The three matrices $\tilde\tau_k=\tau_2\tau_k$ are symmetric, such that $C_\pm\otimes \tau_2\tau_k$ is antisymmetric, as required by the Pauli principle. An $SO(4,{\mathbbm C})$ invariant hyperloop can be written as a product of six factors $\tilde \h(\tilde z)$, with $\tilde z$ belonging to the hypercube $\tilde y$ and obeying eq. \eqref{V1}. We will take all six positions to be different. Furthermore, we will take three factors $\tilde \h_+$ and three factors $\tilde \h_-$ in order to realize the global symmetries of the continuum limit.

\section{Lattice action}
\label{Lattice action}

The lattice action is a sum of local terms ${\cal L}(y)$ for all hypercubes $\tilde y$, where each ${\cal L}(y)$ is a combination of hyperloops. Using only the bilinears \eqref{L4} the local Lorentz symmetry is guaranteed. We need a lattice implementation of the contraction of four derivatives with $\epsilon^{\mu_1\mu_2\mu_3\mu_4}$ in order to realize diffeomorphism symmetry in the continuum limit. As basic building blocks we define 
\ba\label{N4}
{\cal F}^\pm_{\mu\nu}\y&=&\frac{1}{12}\epsilon^{klm}\bar\h^\pm_k\y\nn\\
&\times&\big[\h^\pm_l(\tilde y+w_\mu)-\h^\pm_l(\tilde y-w_\mu)\big]\nn\\
&\times&\big[\h^\pm_m(\tilde y+w_\nu)-\h^\pm_m(\tilde y-w_\nu)\big],
\ea
with $\bar\h\y$ the cell average
\be\label{XXXAX}
\bar\h^\pm_k(\tilde y)=\frac18\sum_\nu\big(\h^\pm_k(\tilde y+w_\nu)+\h^\pm_k(\tilde y-w_\nu)\big)
\ee
A lattice diffeomorphism invariant action in four dimensions can be written as
\be\label{N5}
S=\frac{\alpha}{128}\sum_{\tilde y}\epsilon^{\mu\nu\rho\sigma}{\cal F}^+_{\mu\nu}{\cal F}^-_{\rho\sigma}+c.c.
\ee
We observe that the action is invariant under $\pi/2$-rotations in all six planes spanned by pairs of two coordinates $z^\mu$. It is also odd under all four reflections of a single coordinate, $z^\mu\to -z^\mu$, as well as under diagonal reflections $z^\mu\leftrightarrow z^\nu$ or $z^\mu\to -z^\nu(\mu\neq \nu)$. 

Finally, we note that the three components $\tilde \h^k_+$ in eq. \eqref{L4} transform as a three-component vector with respect to global $SU(2,{\mathbbm C})_L$ gauge transformations. Thus the contraction \eqref{N4} with the invariant tensor $\epsilon^{klm}$ yields a $SU(2,{\mathbbm C})_L$-singlet, and $\f^+_{\mu\nu}(\tilde y)$ is invariant under global $SU(2,{\mathbbm C})_L$ transformations. The lattice action is invariant under global $SU(2,{\mathbbm C})_L\times SU(2,{\mathbbm C})_R$ gauge transformations. It is, however, not invariant under local gauge transformations of this kind. Local gauge transformations transform the factors $\tilde\h^k_\pm$ at different positions $\x j$ differently. If we would like to realize local $SU(2)$ gauge symmetry we would have to replace $(\tilde \tau_k)^{ab}$ in eq. \eqref{L4} by the invariant $\tilde\tau_0=\tau_2$. This is not compatible with local Lorentz symmetry. The $4\times 4$ matrices $C_\pm\otimes \tilde\tau_0$ are symmetric, such that $\tilde \h$ would vanish due to the Pauli principle. One could try to realize a local $U(1)$-symmetry by employing a different structure where only $\tilde \h^3_\pm$ appears. This is, however, not compatible with the required transformation properties of the lattice action with respect to reflections. 

We define the lattice derivatives by the four relations
\be\label{N6}
\h(\tilde y+w_\nu)-\h(\tilde y-w_\nu)=
(x^+_\nu-x^-_\nu)^\mu\hat\partial_\mu\h\y,
\ee
where $x^\pm_\nu=x_{p}(\tilde y\pm w_\nu)$. Here we extend our discussion to a general assignment of points in a manifold $x_p(\tilde z)$ for any discrete label $\tilde z$ of the lattice points. Our special case of a regular lattice corresponds to $x_p(\tilde z)=\Delta\tilde z$. With $\Delta_\nu=(x^+_\nu- x^-_\nu)/2$ the cell volume amounts to 
\ba\label{N7}
V\y&=&2\epsilon_{\mu\nu\rho\sigma}\Delta^\mu_0\Delta^\nu_1\Delta^\rho_2\Delta^\sigma_3\nn\\
&=&\frac{1}{12}
\epsilon_{\mu\nu\rho\sigma}\epsilon^{\mu'\nu'\rho'\sigma'}\Delta^\mu_{\mu'}\Delta^\nu_{\nu'}\Delta^\rho_{\rho'}
\Delta^\sigma_{\sigma'}.
\ea
The volume depends on the particular choice of positions $x_p(\tilde z)$. (We only consider $V(\tilde y)>0$.) Also the expressions for the lattice derivatives $\hat\partial_\mu\h$, which follow from solving eq. \eqref{N6}, depend on this ``positioning of the lattice points''.

Using $\int d^4x=\sum_{\tilde y}V\y$ one finds that the action does not depend on the positioning of the lattice points, if it is expressed in terms of lattice derivatives and a continuous integral $\int d^4 x$, i.e.
\be\label{N8}
S=\frac{\alpha}{16}\int d^4x~\epsilon^{\mu\nu\rho\sigma}\hat{\cal F}^+_{\mu\nu}\hat{\cal F}^-_{\rho\sigma}+c.c.,
\ee
with 
\be\label{N9}
\hat{\cal F}^\pm_{\mu\nu}\y=\frac{1}{12}\epsilon^{klm}\bar\h^\pm_k\y\hat\partial_\mu\h^\pm_l\y\hat\partial_\nu\h^\pm_m\y.
\ee
The positioning dependence of the derivatives is canceled by the one of the volume. This will be crucial for the lattice diffeomorphism symmetry discussed in the next section. 

The continuum limit $\bar \h\to\h,\hat\partial_\mu\to\partial_\mu$, is diffeomorphism invariant due to the contraction of the partial derivatives with the $\epsilon$-tensor. It obtains formally as $\Delta\to 0$ at fixed $y^\mu$ - for details see ref. \cite{CWLD}. We use eq. \eqref{32E} and find for the continuum limit
\be\label{N12}
\hat{\cal F}^\pm_{\mu\nu}\to\pm 4iF^\pm_{\mu\nu}.
\ee
One recovers the diffeomorphism symmetric action \eqref{32F}.

\section{Lattice diffeomorphism invariance}
\label{Lattice diffeomorphism invariance}

In the third part of this work we discuss the lattice equivalent of diffeomorphism symmetry of the continuum action. This ``lattice diffeomorphism invariance'' should be a special property of the lattice action that guarantees diffeomorphism symmetry for the continuum limit and the quantum effective action. We have no fundamental metric or vierbein at our disposal. Neither do we employ geometrical objects as simplices in order to to perform a ``functional integration over geometries.'' Our concept of lattice diffeomorphism invariance differs therefore substantially from the approach in Regge gravity \cite{DIR}. The lattice points are associated to points in a coordinate manifold. The latter is simply a region in ${\mathbbm R}^d$ and we have to formulate an invariance principle for this type of setting. 

In the continuum, the invariance of the action under general coordinate transformations states that it should not matter if fields are placed at a point $x$ or some neighboring point $x+\xi$, provided that all fields are transformed simultaneously according to suitable rules. In particular, scalar fields ${\cal H}(x)$ are simply replaced by ${\cal H}(x+\xi)$. After an infinitesimal transformation the new scalar field $\h '(x)$ at a given position $x$ is related to the original scalar field $\h(x)$ by 
\be\label{1}
\h'(x)=\h(x-\xi)=\h(x)-\xi^\mu\partial_\mu\h(x).
\ee
Diffeomorphism symmetry states that the action is the same for $\h(x)$ and $\h'(x)$. Implicitly the general coordinate transformations assume that the same rule for forming derivatives is used before and after the transformation.

We want to implement a similar principle for a lattice formulation. For this purpose we associate the abstract lattice points $\tilde z=(\zz0,\zz1,\zz2,\zz3)$, with integer $\zz\mu$, with points on a manifold. We consider here a piece of ${\mathbbm R}^d$ with cartesian coordinates $x^\mu=(x^0,x^1,\dots x^{d-1})$, but we do not specify any metric a priori, nor assume its existence. A map $\tilde z\to x^\mu_p(\tilde z)$ defines the positioning of lattice points in the manifold. We can now compare two different positionings, as a regular lattice $x_p^\mu(\tilde z)=\tilde z^\mu\Delta$, or some irregular one with different coordinates $x'^{\mu}_p(\tilde z)$, for the same abstract lattice points $\tilde z$. In particular, we can compare two positionings related to each other by an arbitrary infinitesimal shift $x'^\mu_p=x^\mu_p+\xi^\mu_p(x)$. The notion of an {\em infinitesimal} neighborhood requires a continuous manifold and cannot be formulated for the discrete abstract lattice points $\tilde z$. 

Positioning of the lattice points on a manifold is also required for the notion of a lattice derivative. One can define the meaning of two neighboring lattice points $\tilde z_1$ and $\tilde z_2$ in an abstract sense. A lattice derivative of a field will then be connected to the difference between field values at neighboring sites, $\h(\z1)-\h(\z2)$. For the definition of a lattice derivative $\hat\partial_\mu\h$ we need, in addition, some quantity with dimension of length. This is provided by the positioning on the manifold and follows from solving eq. \eqref{N6} for $\hat\partial_\mu\h$. Furthermore, the positioning of $\tilde z$ on a manifold is a crucial ingredient for the formulation of a continuum limit, where one switches from $\h(\tilde z)$ to $\h(x)$ and derivatives thereof.

If the lattice action is originally formulated in terms of $\h(\tilde z)$ only, its expression in terms of lattice derivatives will in general depend on the positioning, since the relation between $\h(\tilde z)$ and lattice derivatives \eqref{N6} depends on the positioning. We can now state the principle of ``lattice diffeomorphism invariance''. A lattice action is lattice diffeomorphism invariant if its expression in terms of lattice derivatives and a continuous integral does not depend on the positioning of the lattice points. For infinitesimally close positionings the lattice action is then independent of $\xi_p$. The lattice action \eqref{N8} exhibits this property of lattice diffeomorphism invariance. 

The usual discussion of lattice theories considers implicitly a given fixed positioning, for example a regular lattice. We investigate here a much wider class of positionings. Only the comparison of different positionings allows the formulation of lattice diffeomorphism invariance. One can show that the continuum limit of a lattice diffeomorphism invariant action exhibits diffeomorphism symmetry \cite{CWLD}. Also the quantum effective action is invariant under general coordinate transformations. This extends to the effective action for the metric which appears in our setting as the expectation value of a suitable collective field. The gravitational field equations are therefore covariant, with a similar general structure as in general relativity. 

In order to show diffeomorphism symmetry of the continuum limit and the effective action a central ingredient is the observation that diffeomorphism transformations can be realized by repositionings of the lattice variables, without transforming the lattice variables themselves. One employs the concept of interpolating functions \cite{CWLD} and defines a version of partial derivatives of interpolating functions that takes into account the lack of knowledge of details of the interpolation. At the positions of lattice cells these derivatives equal the lattice derivatives. For smooth interpolating fields they coincide with the standard definition of partial derivatives. In this view, the lattice does not reflect a basic discreteness of space and time. It rather expresses the fact that only a finite amount of information is available in practice, and that arbitrarily accurate continuous functions are an idealization since they require an infinite amount of information. In a sense, we treat continuous functions similar to numerical simulations. In our formulation diffeomorphism transformations are nothing else than moving the lattice points, where the information about the function is given, within a manifold. Diffeomorphism symmetry is realized if the action in terms of fields and their derivatives does not notice this change in positions. 

\section{Lattice diffeomorphism invariance in two dimensions}
\label{Lattice diffeomorphism invariance in two dimensions}

Basic construction principles of a lattice diffeomorphism invariant action can be understood in two dimensions. We label abstract lattice points by two integers $\tilde z=(\tilde z^0,\tilde z^1)$, with $\tilde z^0+\tilde z^1$ odd. For the discussion of lattice diffeomorphism invariance only the transformation of $\h_k$ as a scalar matters. Our discussion therefore also applies for fundamental scalars $\h_k$ \cite{CWLD}. For lattice spinor gravity $\h_k$ is again a fermion bilinear. We use for every lattice point two species, $a=1,2$, of two-component complex Grassmann variables $\varphi^a_\alpha(\tilde z),\alpha=1,2$, or equivalently eight real Grassmann variables $\psi^a_\gamma(\tilde z),\gamma=1\dots 4$, with $\varphi^a_1(\tilde z)=\psi^a_1(\tilde z)+i\psi^a_3(\tilde z)~,~\varphi^a_2(\tilde z)=\psi^a_2(\tilde z)+i\psi^a_4(\tilde z)$. The functional measure \eqref{1A} is replaced by
\be\label{x2}
\int \D\psi=\prod_{\tilde z}\prod_\gamma\big (d\psi^1_\gamma(\tilde z)d\psi^2_\gamma(\tilde z)\big).
\ee

We introduce the bilinears $\h_k$ as in eq. \eqref{L4}, with $\alpha,\beta=1,2$,
and define the action as a sum over local cells located at $\tilde y=(\tilde y^0,\tilde y^1)$, with $\tilde y^\mu$ integer and $\tilde y^0+\tilde y^1$ even, as in eq. \eqref{L1}. Each cell consists of four lattice points that are nearest neighbors of $\tilde y$, denoted by $\tilde x_j(\tilde y),j=1\dots 4$. Their lattice coordinates are 
$\tilde z\big(\tilde x_1(\tilde y)\big)=(\tilde y^0-1,\tilde y^1)~,~\tilde z\big(\tilde x_2(\tilde y)\big)=(\tilde y^0,\tilde y^1-1)~,~\tilde z\big(\tilde x_3(\tilde y)\big)=(\tilde y^0,\tilde y^1+1)$, and $\tilde z\big(\tilde x_4(\tilde y)\big)=(\tilde y^0+1,\tilde y^1)$. The local term $\cl(\tilde y)$ involves lattice fields on the four sites of the cell that we denote by $\h_k(\tilde x_j)$. We choose 
\ba\label{6}
\cl(\tilde y)&=&\frac{1}{48}\epsilon^{klm}\big[\h_k(\tilde x_1)+\h_k(\tilde x_2)+\h_k(\tilde x_3)+\h_k(\x4)\big]\nn\\
&&\times\big[\h_l(\x4)-\h_l(\x1)\big]
[\h_m(\x3)-\h_m(\x2)\big].
\ea
At this point no notion of a manifold is introduced. We specify only the connectivity of the lattice by grouping lattice points $\tilde z$ into cells $\tilde y$ such that each cell has four points and each point belongs to four cells. This defines neighboring cells as those that have two common lattice points. Neighboring lattice points belong to at least one common cell.

We now proceed to an (almost) arbitrary positioning of the lattice points on a piece of ${\mathbbm R}^2$ by specifying positions $x^\mu_p(\tilde z)$. This associates to each cell a ``volume'' $V(\tilde y)$,
\be\label{7}
V(\tilde y)=\frac12\epsilon_{\mu\nu}(x^\mu_4-x^\mu_1)(x^\nu_3-x^\nu_2),
\ee
with $\epsilon_{01}=-\epsilon_{10}=1$ and $x^\mu_j$ shorthands for the positions of the sites $\tilde x_j$ of the cell $\tilde y$, i.e. $x^\mu_j=x^\mu_p\Big(\tilde z\big(\x{j}(\tilde y)\big)\Big)$. The volume corresponds to the surface inclosed by straight lines joining the four lattice points $\x j(\tilde y)$ of the cell in the order $\x1,\x2,\x4,\x3$. For simplicity we restrict the discussion to ``deformations'' of the regular lattice, $x^\mu_p=\tilde z^\mu\Delta$, where $V(\tilde y)$ remains always positive and the path of one point during the deformation never touches another point or a straight line between two other points at the boundary of the surface. We use the volume $V(\tilde y)$ for the definition of an integral over the relevant region of the manifold
\be\label{8}
\int d^2 x=\sum_{\tilde y}V(\tilde y),
\ee
where we define the region by the surface covered by the cells appearing in the sum.

We next express the action \eqref{L1}, \eqref{6} in terms of average fields in the cell
\be\label{9}
\h_k(\tilde y)=\frac14\sum_j\h_k\big(\x{j}(\tilde y)\big)
\ee
and lattice derivatives associated to the cell
\ba\label{9A}
\hat\partial_0\h_k\y&=&\frac{1}{2V\y}\Big\{(x^1_3-x^1_2)\big(\h_k(\x4)-\h_k(\x1)\big)\nn\\
&&-(x^1_4-x^1_1)\big(\h_k(\x3)-\h_k(\x2)\big)\Big\},\nn\\
\hat\partial_1\h_k\y&=&\frac{1}{2V\y}\Big\{(x^0_4-x^0_1)\big(\h_k(\x3)-\h_k(\x2)\big)\nn\\
&&-(x^0_3-x^0_2)\big(\h_k(\x4)-\h_k(\x1)\big)\Big\}.
\ea
For the pairs $(\x{j_1},\x{j_2})=(\x4,\x1)$ and $(\x3,\x2)$ the lattice derivatives obey 
\be\label{10}
\h_k(\x{j_1})-\h_k(\x{j_2})=
(x^\mu_{j_1}-x^\mu_{j_2})\hat\partial_\mu\h_k(\tilde y),
\ee
similar to eq. \eqref{N6}. In terms of average and derivatives all quantities in $\cl(\tilde y)$ depend on the cell variable $\tilde y$ or the associated position of the cell $x^\mu_p(\tilde y)$ that we take somewhere inside the surface of the cell, the precise assignment being unimportant at this stage. In this form we denote $\cl\y$ by $\hat\cl(\tilde y;x_p)$ or $\hat\cl(x;x_p)$, where $\hl(x)$ only depends on quantities with support on discrete points in the manifold corresponding to the cell positions. We indicate explicitly the dependence on the choice of the positioning by the argument $x_p$. 

The action appears now in a form referring to the positions on the manifold
\be\label{11}
S(x_p)=\tilde \alpha\int d^2x\bar\cl(\tilde y;x_p)+c.c=\tilde\alpha\int d^2x\bar\cl(x;x_p)+c.c.,
\ee
with 
\be\label{12}
\bar\cl(\tilde y;x_p)=\bar\cl(x;x_p)=\frac{\hat\cl(\tilde y;x_p)}{V(\tilde y;x_p)}.
\ee
Lattice diffeomorphism invariance states that for fixed $\h(\tilde y)$ and $\hat\partial_\mu\h(\tilde y)$ the ratio $\bar\cl(\tilde y;x_p)$ is independent of the positioning, or independent of $\xi_p$ for infinitesimal changes of positions $x'_p=x_p+\xi_p$,
\be\label{12A}
\bar\cl(\tilde y;x_p+\xi_p)=\bar\cl(\tilde y;x_p)~,~S(x_p+\xi_p)=S(x_p).
\ee

The $\xi_p$-independence of $\bar\cl(\tilde y;x_p)$ means that the dependence of $V(\tilde y;x_p)$ and $\hat\cl(\tilde y;x_p)$ on the positioning $x_p$ must cancel. Inserting eqs.~\eqref{9},~\eqref{10} in eq. \eqref{6} yields
\be\label{13}
\hat\cl\y=\frac{1}{12}\epsilon^{klm}V\y\h_k\y\epsilon^{\mu\nu}\hat\partial_\mu\h_l\y\hat\partial_\nu\h_m\y,
\ee
and we find indeed that the factor $V\y$ cancels in $\bar\cl\y=\hl\y/V\y$. Thus the action \eqref{L1}, \eqref{6} is  lattice diffeomorphism invariant. This property is specific for a certain class of actions - for example adding to $\epsilon^{klm}$ a quantity $s^{klm}$ which is symmetric in $l\leftrightarrow m$ would destroy lattice diffeomorphism symmetry. For all typical lattice theories the formulation of $\cl\y$ only in terms of next neighbors and common cells (not using a distance) does not refer to any particular positioning. However, once one proceeds to a positioning of the lattice points and introduces the concept of lattice derivatives, the independence on the positioning of $\bar\cl(\tilde y;x_p)$ for fixed $\h\y$ and $\hat\partial_\mu\h\y)$ is not shared by many known lattice theories. For example, standard lattice gauge theories are not lattice diffeomorphism invariant. 

Using the concept interpolating functions for fermion-bilinears \cite{CWLD} the continuum limit obtains by replacing lattice derivatives by partial derivatives and all average fields by local fields. This yields for the continuum action as a functional of the interpolating fields
\be\label{x1}
S=\frac{\tilde\alpha}{12}\int d^2x\epsilon^{klm}\epsilon^{\mu\nu}
\h_k(x)\partial_\mu\h_l(x)\partial_\nu\h_m(x)+c.c.
\ee
The lattice derivatives for the Grassmann variables are defined similar to eq. \eqref{10} by the two relations
\ba\label{x4}
\varphi^a_\alpha(\x{j_1})-\tilde\varphi^a_\alpha(\x{j_2})=(x^\mu_{j_1}-x^\mu_{j_2})\hat\partial_\mu\varphi^a_\alpha\y
\ea
for $(j_1,j_2)=(4,1)$ and $(3,2)$. With 
\ba\label{x5}
&&\h_k(\x{j_1})-\h_k(\x{j_2})=
\big(\varphi^a_\alpha(\x{j_1})
+\varphi^a_\alpha(\x{j_2})\big)
(\tau_2)_{\alpha\beta}\nn\\
&&\qquad \qquad\times(\tau_2\tau_k)^{ab}\big(\varphi^b_\beta(\x{j_1})-\varphi^b_\beta(\x{j_2})\big),
\ea
and using reordering of the Grassmann variables, one obtains from eq. \eqref{6}
\ba\label{x6}
{\cal L}(y)&=&-8 i\alpha A\y\big(\varphi^a_\alpha(\x4)-\varphi^a_\alpha(\x1)\big)
(\tau_2)_{\alpha\beta}(\tau_2)^{ab}\nn\\
&&\times\big (\varphi^b_\beta(\x3)-\varphi^b_\beta(\x2)\big)+\dots,
\ea
with 
\be\label{x7}
A\y=\bar\varphi^1_1\y\bar\varphi^1_2\y\bar\varphi^2_1\y\bar\varphi^2_2\y,
\ee
and $\bar\varphi^a_\alpha\y$ the cell average. The dots indicate terms that do not contribute in the continuum limit. In terms of lattice derivatives \eqref{x4} one finds the action
\be\label{x8}
\bar\cl\y=-8i\tilde \alpha A\y\epsilon^{\mu\nu}\hat\partial_\mu\varphi^a_\alpha\y(\tau_2)_{\alpha\beta}(\tau_2)^{ab}\hat\partial_\nu
\varphi^b_\beta\y+\dots
\ee
For fixed spinor lattice derivatives \eqref{x4} the leading term \eqref{x8} is again lattice diffeomorphism invariant.

The continuum limit \eqref{x1} can be expressed in terms of spinors using $\partial_\mu\h_k(x)=2\varphi(x)\tau_2\otimes\tau_2\tau_k\partial_\mu\varphi(x)$, where the first $2\times 2$ matrix $E$ in $E\otimes F$ acts on spinor indices $\alpha$, the second $F$ on flavor indices $a$.   With
\be\label{N1}
F_{\mu\nu}=-A\partial_\mu\varphi\tau_2\otimes\tau_2\partial_\nu\varphi
\ee
one obtains
\be\label{N2}
S=4i\tilde \alpha\int d^2x\epsilon^{\mu\nu}F_{\mu\nu}+c.c.,
\ee
in accordance with eq. \eqref{x8}. Two comments are in order: (i) For obtaining a diffeomorphism invariant continuum action it is sufficient that the lattice action is lattice diffeomorphism invariant up to terms that vanish in the continuum limit. (ii) The definition of lattice diffeomorphism invariance is not unique, differing, for example, if we take fixed lattice derivatives \eqref{9A} for spinor bilinears or the ones \eqref{x4} for spinors. It is sufficient that the action is lattice diffeomorphism invariant for {\em one} of the possible definitions of lattice derivatives kept fixed. 

We finally note that $A$ and $F_{\mu\nu}$ are invariant under $SO(4,{\mathbbm C})$ transformations. This symmetry group rotates among the four complex spinors $\varphi^a_\alpha$, with complex infinitesimal rotation coefficients. For real coefficients, one has $SO(4)$, whereas other signatures as $SO(1,3)$ are realized if some coefficients are imaginary. The continuum action \eqref{N2} or \eqref{x1} exhibits a local $SO(4,{\mathbbm C})$ gauge symmetry. A subgroup of $SO(4,{\mathbbm C})$ is the two-dimensional Lorentz group $SO(1,1)$. The action \eqref{6} is therefore a realization of lattice spinor gravity \cite{LD4} in two dimensions. 

The extension of this discussion to four dimensions is straightforward. One verifies that the lattice action \eqref{N5} is lattice diffeomorphism invariant. One can also define the concept of lattice diffeomorphism transformations \cite{CWLD} which is directly linked to the repositioning of lattice points within a continuous manifold. 

\section{Effective action} 
\label{Effective action}

The quantum effective action for fermions is introduced in the usual way by introducing Grassmann valued sources and making a Legendre transform of $\ln Z$. We can also introduce the effective action for bosonic collective fields. As an example, we discuss here first the fermion bilinear $\h_k$. 

The generating functional for the connected Greens functions of collective lattice variables $\h_k$ is defined in the usual way 
\be\label{22O}
W\big[J\y\big]=\ln\int{\cal D}\psi\h\exp\big\{-S+\sum_{\tilde y}\big(\h_k\y J^*_k\y+c.c\big)\big\},
\ee
with 
\be\label{22P}
\frac{\delta W}{\delta J^*_k\y}=\kl\h_k\y\kr=h_k\y.
\ee
(We don not write explicitly the boundary terms $g_f$ and $g_{in}$ in eq. \eqref{1A} for $Z$. They  may be incorporated formally into $\int {\cal D}\psi$.) In the continuum limit the source term becomes 
\be\label{22Q}
\sum_{\tilde y}\h_k\y J^*_k\y=\int_x\h_k\y j^*_k\y=\int_x\h_k(x)j^*_k(x),
\ee
where the lattice source field $j\y=J\y/V\y$ transforms as a scalar density under lattice diffeomorphisms. One also may define 
\be\label{22R}
\Gamma[h,J]=-W[J]+\sum_{\tilde y}\big(h_k\y J^*_k\y+c.c.\big),
\ee
which becomes the usual quantum effective action $\Gamma[h]$ (generating functional of $1$PI-Greens functions) if we solve eq. \eqref{22P} for $J^*\y$ as a functional of $h\y$ and insert this solution into eq. \eqref{22R}. 

We have shown in ref. \cite{CWLD} that the effective action $\Gamma[h]$ is lattice diffeomorphism invariant. Its continuum limit exhibits the usual diffeomorphism symmetry if $h(x)$ transforms as a scalar. The proof relies on the observation that if the action does not ``notice'' the positioning of lattice points on the coordinate manifold, the same holds true for the effective action. No information about a specific positioning is introduced by the construction \eqref{22O}-\eqref{22R}.

\section{Metric}
\label{Metric}

In the fourth part of this note we discuss the emergence of geometry from our formulation of lattice spinor gravity. So far we have used the coordinates $x^\mu$ only for the parametrization of a region of a continuous manifold. We have not used the notion of a metric and the associated ``physical distance''. (The physical distance differs from the coordinate distance $|x-y|$, except for the metric $g\mn=\delta\mn$.) The notion of a metric and the associated physical distance, topology and geometry can be inferred from the behavior of suitable correlation functions \cite{CWG}. Roughly speaking, for a euclidean setting the distance between two points $x$ and $y$ gets larger if a suitable properly normalized connected two-point function $G(x,y)$ gets smaller. This is how one world ``measure'' distances intuitively. In our case we may consider the two point function for collective fields $G(x,y)=\kl \h_k(x)\h_k(y)\kr$. 

We define the metric as
\ba\label{M2}
g\mn(x)&=&\frac12(\kl G\mn(x)+G^*\mn(x)\kr),\nn\\
G\mn(x)&=&\mu^{-2}_0\sum_k\partial_\mu\h_k(x)\partial_\nu\h_k(x).
\ea
Here $\h_k(x)$ stands for the continuum limit or for a suitable interpolating field. The real normalization constant $\mu^{-1}_0$ has dimension of length such that $G\mn$ and $g\mn$ are dimensionless. In general, the elements $\kl G_{\mu\nu}(x)\kr$ can be complex such that $g_{\mu\mu}$ can be positive or negative real numbers. The signature of the metric is not defined a priori. Points where $\det\big(g\mn(x)\big)=0$ indicate singularities - either true singularities or coordinate singularities. More generally, the geometry and topology (e.g. singularities, identification of points etc.) of the space can be constructed from the metric \cite{CWG}. The metric is the central object in general relativity and appears in our setting as the expectation value of a suitable collective field. 

On the lattice we may use interpolating functions \cite{CWLD} for $\h_k(x)$. For $x$ coinciding with the position of one of the cells $y_n=x_p(\tilde y_n)$ the derivative $\partial_\mu\h_k(x)$ is then given by the lattice derivative $\hat\partial_\mu\h_k(\tilde y)$. For these values of $x$ the field $G\mn(x)$ can be expressed by lattice quantities
\ba\label{GA}
G\mn(x)&=&\mu^{-2}_0\sum_k\hat\partial_\mu\h_k(\tilde y)\hat\partial_\nu\h_k(\tilde y)\nn\\
&=&\mu^{-2}_0a^{\tilde\mu}_\mu(x)a^{\tilde\nu}_\nu(x)G^{(L)}_{\tilde\mu\tilde\nu},
\ea
with ``lattice metric''
\be\label{M3}
G^{(L)}_{\tilde\mu\tilde\nu}=p_{k,\tilde\mu}p_{k,\tilde\nu}
\ee
and 
\be\label{M4}
p_{k,\tilde\mu}=\h_k(\tilde y+v_{\tilde\mu})-\h_k(\tilde y-v_{\tilde\mu}).
\ee
Similar to the lattice derivatives, the $x$-dependence of the metric arises only through the functions $a^{\tilde\mu}_\mu(x)$ which reflect the positioning of the lattice points. These functions obey
\be\label{78A}
\hat\partial_\mu\h_k(\tilde y)=a^{\tilde\mu}_\mu(x)p_{k,\tilde\mu},
\ee
and their explicit form can  be extracted from eq. \eqref{12}. For interpolating functions $\h_k(x)$ transforming as scalars under general coordinate transformations the metric \eqref{M2} transforms as a covariant second rank symmetric tensor. This is matched by the transformation properties of the expression \eqref{GA} under lattice diffeomorphisms. 

As a particular positioning we can use the regular lattice $x^\mu(\tilde z)=\Delta\tilde z^\mu$. This corresponds to a fixed choice of coordinates in general relativity. With this choice one has $d^{\tilde\mu}_\mu=2\Delta\delta^{\tilde\mu}_\mu,V\y=2\Delta^2$ and therefore
\be\label{M5}
a^{\tilde\mu}_\mu(x)=\frac{1}{2\Delta}\delta^{\tilde\mu}_\mu.
\ee 
Choosing $\mu^{-2}_0=4\Delta^2$, the collective field $G\mn$ in eq. \eqref{GA} coincides with the lattice metric $G^{(L)}\mn$ in eq. \eqref{M3}.

Properties of the metric can often be extracted from symmetries. If the expectation values preserve lattice translation symmetry the metric $g\mn(x)$ will be independent of $x$. Invariance under a parity reflection implies $g_{0k}=g_{k0}=0$. Symmetry of the expectation values under lattice rotations would imply a flat euclidean metric $g_{\mu\nu}\sim \delta_{\mu\nu}$. A Minkowski metric $g\mn=\eta\mn$ requires that the expectation values violate the euclidean rotation symmetry.  

\section{Effective action for gravity and gravitational field equations}
\label{Effective action for gravity and gravitational field equations}

The quantum effective action for the metric, $\Gamma[g_{\mu\nu}]$, can be constructed in the usual way by introducing sources for the collective field,
\ba\label{MB}
W[\tilde T]&=&\ln \int {\cal D}\h\exp\big\{-S+\int_xG^R\mn(x)\tilde T^{\mu\nu}(x)\big\},\nn\\
G^R\mn&=&\frac12(G\mn+G^*\mn)\quad,\quad\frac{\delta W[\tilde T]}{\delta\tilde T^{\mu\nu}(x)}=g\mn(x).
\ea
Solving formally for $\tilde T^{\mu\nu}$ as a functional of $g\mn$, the quantum effective action for the metric obtains by a Legendre transform
\be\label{MC}
\Gamma[g\mn]=-W+\int_x g\mn(x)\tilde T^{\mu\nu}(x). 
\ee
The metric obeys the exact quantum field equation
\be\label{MD}
\frac{\delta\Gamma}{\delta g\mn(x)}=\tilde T\mn(x),
\ee
and we realize that $\tilde T^{\mu\nu}$ can be associated to the energy momentum tensor $T^{\mu\nu}$ by $\tilde T^{\mu\nu}=\sqrt{g}T^{\mu\nu}~,~g=|\det g\mn|$.

Under a general coordinate transformation $\h_k(x)$ transforms as a scalar
\be\label{62}
\delta \h_k(x)=-\xi^\nu\partial_\nu \h_k(x).
\ee
This implies that $\partial_\mu \h_k$ and $G^R\mn$ transform as covariant vectors and second rank symmetric tensors, respectively. In consequence, $\tilde T^{\mu\nu}$ transforms as a contravariant tensor density, with $T^{\mu\nu}$ a symmetric second rank tensor. Thus $\int_xG^R\mn\tilde T^{\mu\nu}$ and $\int_x g\mn\tilde T^{\mu\nu}$ are diffeomorphism invariant, and $\Gamma[g\mn]$ is diffeomorphism invariant if $W[\tilde T]$ is diffeomorphism invariant. This is indeed the case for $\tilde T\mn$ transforming as a tensor density \cite{CWLD} - the argument is similar as for the diffeomorphism symmetry of $\Gamma\big[h(x)\big]$. 

The functional integral \eqref{MB} is well defined and regularized for a finite number of lattice points. Therefore also $\Gamma\big[g\mn(x)\big]$ is a well defined functional that is, in principle, unambiguously calculable. (More precisely, this holds for all metrics for which the third equation \eqref{MB} is invertible.)

A key question concerns the general form of the effective action $\Gamma[g\mn]$. If $\Gamma$ is diffeomorphism invariant and sufficiently local in the sense that an expansion in derivatives of $g\mn$ yields a good approximation for slowly varying metrics, then only a limited number of invariants as a cosmological constant or Einstein's curvature scalar $R$ contribute at long distances. The signature of the metric is not fixed a priori. For $g\neq 0$ the inverse metric $g^{\mu\nu}$ is well defined - this contrasts with the Grassmann element $G\mn$ or $G^R\mn$ for which no inverse exists. The existence of $g^{\mu\nu}$ opens the possibility that $\Gamma[g\mn]$ also involves the inverse metric. Two dimensions are special for gravity since the graviton does not propagate. Our construction generalizes, however, in a straightforward way to four dimensions.

\section{Conclusions and discussion}
\label{Conclusions}

We have constructed a lattice regularized functional integral for fermions with local Lorentz symmetry. The continuum limit of both the action and the quantum effective action exhibits invariance under general coordinate transformations. We thus have realized the first four of the six criteria for realistic quantum gravity that we have specified in the introduction. The remaining two criteria \eqref{AA1} and \eqref{A1a} depend on the form of the quantum effective action for the metric. The diffeomorphism invariance of the effective action suggests that it can describe a massless graviton if the cosmological constant vanishes. For a verification of this conjecture one needs, however, an explicit computation of the long wavelength limit of the effective action.

The symmetry properties of our model suggest that it can be used as a promising starting point for realistic quantum gravity. We have only sketched the way to geometry. Much remains to be done before the effective action for the composite metric can be computed explicitly. For our regularized model this issue is, at least, well defined. However, only an explicit calculation can settle the issue if diffeomorphism invariant terms involving explicit length scales, as a cosmological constant or Einstein's curvature scalar multiplied by the Planck mass, can be generated by fluctuations. The classical continuum action \eqref{30F} is dilatation symmetric - the only coupling $\alpha$ is dimensionless. If the effective action for the graviton preserves dilatation symmetry no dimensional couplings can be present. In this case one would expect gravitational invariants involving two powers of the curvature tensor, as $R_{\mu\nu\rho\sigma}R^{\mu\nu\rho\sigma}$, $R_{\mu\nu}R^{\mu\nu}$ or $R^2$. Also composite scalar fields may play a role, such that terms $\sim \xi^2 R$ can induce an Einstein-Hilbert term in the effective action by spontaneous dilatation symmetry breaking through an expectation value of $\xi$ \cite{CWDil}, \cite{Fu}. As an alternative, an explicit mass scale could be generated by running couplings, which constitute a dilatation anomaly through quantum fluctuations. 

For a regularized functional integral realizing lattice diffeomorphism invariance the lattice distance $\Delta$ does not introduce an explicit length scale. It neither appears in the lattice action nor in the continuum limit of the action. The parameter $\Delta$ only characterizes a particular regular positioning of the abstract lattice points on a manifold, and one can vary its value by repositioning. This absence of a length scale suggests that the ultraviolet limit of quantum gravity is characterized by an ultraviolet fixed point. Such a fixed point would realize the ``asymptotic safety'' scenario for non-perturbative renormalizable gravity \cite{Wei}. Recent progress \cite{Reu}, \cite{Cod} in computations of the flow of gravitational couplings, based on functional renormalization of the effective average action or flowing action \cite{CWRG}, \cite{WR}, give many hints in this direction.

Besides the metric, a consistent coupling of fermions to gravity also needs the vierbein. In our formulation of lattice spinor gravity we have several candidates of the type 
\be\label{ZZA}
\tilde E^m_\mu=\varphi^a C\gamma^m_M\partial_\mu\varphi^b V_{ab},
\ee
with $C=C_1$ or $C_2$ defined in eq. \eqref{W2} and $V_{ab}$ a suitable $2\times 2$ matrix in flavor space. All objects \eqref{ZZA} transform as vectors under general coordinate transformations, and as vectors under global generalized $SO(4,{\mathbbm C})$-Lorentz transformations. (Further objects transforming as vectors under global $SO(1,3)$-transformations can be constructed by replacing $\varphi_\alpha$ by a suitable linear combination of $\varphi^*_\beta$.) From this point of view the expectation value 
\be\label{ZZAa}
e^m_\mu=\kl \tilde E^m_\mu\kr/\mu_e,
\ee
with $\mu_e$ a suitable mass scale, resembles in many aspects the vierbein. 

There are, however, also new unfamiliar features. The bilinear $\tilde E^m_\mu$ does not transform as a vector under local Lorentz transformations, but rather acquires an inhomogeneous piece \cite{HCW,CWSG,CWLL}. By construction the quantum effective action for $e^m_\mu$, which is formulated similarly to the effective action for the collective metric \eqref{MC} or the scalar bilinear $h_k$ \eqref{22R}, is invariant under local Lorentz transformations.  In view of the inhomogeneous transformation property one may expect some differences to Cartan's formulation of gravity \cite{Ca}. 

As another striking feature we observe that $\tilde E^m_\mu$ does not transform as a singlet with respect to the gauge transformations $SU(2)_L\times SU(2)_R$ which act in flavor space. (Exceptions are particular subgroups for particular choices of $V_{ab}$.) This hints to a more intrinsic entanglement between gauge transformations and Lorentz transformations. It remains to be seen if this new form of ``gauge-gravity unification'' could lead to observable effects.

\end{document}